# Bayesian hierarchical space-time models to improve multispecies assessment by combining observations from disparate fish surveys

Chibuzor C. Nnanatu*, Murray S. A. Thompson, Michael A. Spence, Elena Couce, Jeroen van der Kooij and Christopher P. Lynam

*Centre for Environment, Fisheries and Aquaculture Science, Cefas Laboratory, Lowestoft, NR33 OHT, UK*

*Corresponding author: tel: +441502 524541; e-mail: chibuzor.nnanatu@cefas.co.uk

## Abstract

Many wild species affected by human activities require multiple surveys with differing designs to capture behavioural response to wide ranging habitat conditions and map and quantify them. While data from for example intersecting but disparate fish surveys using different gear, are widely available, differences in design and methodology often limit their integration. Novel statistical approaches which can draw on observations from diverse sources could enhance our understanding of multiple species distributions simultaneously and thus provide vital evidence needed to conserve their populations and biodiversity at large. Using a novel Bayesian hierarchical binomial-lognormal hurdle modelling approach, we combined and analysed acoustic and bottom trawl survey data for herring, sprat and northeast Atlantic mackerel in the North Sea. These models were implemented using the Integrated Nested Laplace Approximation (INLA) technique in conjunction with the Stochastic Partial Differential Equation (SPDE) framework. These models accounted for differences in gear efficiencies and inherent spatio-temporal autocorrelations within the data. Models that drew on the combined surveys consistently provided more precise and stable estimates of pelagic species biomass and distribution compared to the individual survey components considered independently, justifying our approach. By accounting for gear-specific efficiencies across surveys in addition to increased spatial coverage (sample size), we gained larger statistical power with greatly minimised uncertainties in estimation. By combining information from disparate fish surveys through advanced statistical modelling, the merits of each survey can be utilised to better assess populations of multiple species simultaneously with a more complete understanding of their distributions including in areas with only a single survey. Our statistical approach provides a methodological development to improve the evidence base for multispecies assessment and marine ecosystem-based management. And on a broader scale, it could be readily applied where disparate biological surveys and sampling methods intersect, e.g. to provide information on biodiversity patterns using global datasets of species distributions.

**Keywords:** Species distribution, pelagic species biomass, acoustic and trawl surveys, gear efficiency, Bayesian hurdle models, spatio-temporal hurdle model, ecosystem-based management.

## Introduction

The distribution of many wild species is not captured by any single survey, so developing statistically robust approaches that incorporate observations from diverse sources is often essential to describe them and broader patterns in biodiversity (Zhou *et al.* 2014; Zipkin *et al.* 2017; Kelley and Sherman, 2018; Moriarty *et al.* 2020). This is particularly acute in marine ecosystems where human activities such as fishing are widespread and the target fish assemblages require observations at large spatial and temporal scales as a prerequisite to understanding species distributions, their population statuses, and biodiversity (Branch et al. 2010).

Many shelf seas are subject to considerable survey effort although sampling methods differ depending on the target species. For example, bottom trawl and acoustic surveys provide complementary but essentially independent snapshots of demersal and pelagic species distributions, respectively. Typically, these are treated in isolation but, if combined, they could be used to better inform stock assessment and ecosystem-based management (Kelley and Sherman, 2018). Herring (*Clupea harengus*), sprat (*sprattus sprattus*), and Northeast Atlantic mackerel (*Scomber scombrus*) found in the North Sea represent such a case study. They are migratory, shoaling, pelagic fish, which act as important intermediate consumers, can be highly abundant and are targeted by some of the region's largest commercial fisheries (Engelhard et al 2014; Lynam et al. 2017). Understanding the impacts of fisheries and environmental change on the distribution and abundance of these species is therefore critical to sustainably manage human impacts on the North Sea ecosystem. Because of their wide distribution, seasonal migrations and different behaviours through the year, no single survey or sampling method captures any, let alone all, of their stocks effectively. Monitoring of small pelagic fish species often uses fisheries acoustic methods (Simmonds and MacLennan, 2005), for example the HERAS survey in the North Sea (ICES, 2020). High resolution (metres and centimetres in horizontal and vertical domain respectively) quantitative information on the biota in the water column is obtained along a series of parallel transects ensuring that patchily distributed aggregations (schools) are sampled, something that cannot be achieved practically with trawl deployments alone. Pelagic trawls are also used during these Acoustic

Surveys (AS) to ground-truth the acoustic marks and collect biological information. The acoustic data are then partitioned by species and age, based on the catch composition of the trawl. Sources of uncertainty unique to AS which can affect the reliability of estimates include: limited capability of detecting fish near the surface or close to the seabed; changes in fish behaviour, for example when fish no longer school and disperse through the water column; varying fish target strength; and variability in the catchability of species using trawl gear during AS to determine species ratios (Kotwicki et al 2017; Walker et al 2017). AS also often follow a systematic transect pattern (e.g., ICES, 2005; van der Kooij et al., 2015) so they are inherently spatially and temporally autocorrelated.

The International Bottom Trawl Survey (IBTS; ICES, 2012; 2019) provides another source of data used in the stock assessment of herring and sprat in the North Sea. Because the trawls used inevitably pass through pelagic habitat during deployment and have relatively high vertical openings, they tend to catch small pelagic species. However, "hit-and-miss" catch-rates during the IBTS caused by pelagic species schooling behaviour can give rise to high uncertainty in their population estimates.

Typical to intersecting but disparate surveys, AS and IBTS each have their own relative unit and sampling biases, because of this they are currently treated as separate indices that are used as independent sources of information for stock assessment. Yet, there is an increasing demand for methods that draw on data from multiple sources within a single analytical framework to improve our understanding of fish distribution and abundance (e.g., Zipkin *et al.* 2017). Species distribution models which integrate different types of observations often do not account for differences between gears (e.g., Guisan et al. 2017; Jarvie and Svenning, 2018; Cheung et al. 2009; Fernandes et al. 2020). Where gear efficiencies have been modelled, studies have incorporated either trawl surveys or a trawl and an acoustic survey. For example, acoustic and trawl surveys have been applied concurrently to improve population assessments of species that have a large vertical distribution during different life stages (Kotwicki et al. 2017); and the efficiency of multiple trawl gears has been estimated to integrate disparate data sources to increase the spatio-temporal scale of fish stock assessments (Walker et al. 2017). Until now, this has not been extended to include multiple disparate multispecies trawl surveys with an acoustic survey in both space and time or applied in a single, fully Bayesian analytical framework which can incorporate multiple sources of uncertainty (Quiroz et al. 2015; Juntunen et al. 2012).

Hierarchical models offer a powerful statistical framework for combining data from multiple sources by explicitly specifying the observation and process models, and allow the incorporation of multiple sources of variations and spatio-temporal autocorrelations within the data through latent parameters (Gelfand, 2010; Sadykova et al. 2017; Justuntun et al. 2012, Pinto et al. 201, Cosandey-Godin. 2015, Quiroz et al. 2015). Moreover, implementing hierarchical models through a robust joint likelihood modelling approach (e.g. Zwolsinki et al. 2009; Moriarty et al. 2020) and within the Bayesian statistical paradigm (Justuntun et al. 2012; Sadykova et al. 2017; Quiroz et al. 2015) enables more accurate spatio-temporal predictions. Specifically, Bayesian hierarchical joint (hurdle) models for zero-inflated data hold promise here because multiple sources of variability are simultaneously accounted for, while at the same time uncertainties in parameter estimations are easily quantified (Sadykova et al. 2015; Quiroz et al. 2015).

Current Bayesian hierarchical models applied to assess fish species distribution and abundance over a continuously indexed spatial surface from acoustic and trawl surveys have mostly relied on simulation-based methods such as Markov chain Monte Carlo (MCMC) techniques (e.g., Juntunen et al 2012; Kotwicki et al. 2017) for parameter estimation. Such methods are well known to be computationally expensive and inefficient (e.g., Sadykova et al. 2017; Quiroz et al. 2015) as the inversion of the implied covariance matrix $\Sigma$ of the continuous Gaussian Process (GP) would require costly computations of order $O(n^3)$. This computational challenge popularly known as the 'big *n* problem' (Jona Lasino et al., 2012) is commonly encountered within ecological studies especially when interest is on elucidating influence of the spatial or spatio-temporal random effects on the outcome based on georeferenced or geostatistical data (Blangiardo et al. 2013; Gelfand *et al*. 2010).

The integrated nested Laplace Approximation (INLA; Rue et al. 2009) in conjunction with the Stochastic Partial Differential equations (SPDE; Lindgren et al., 2011) offer a novel, accurate and computationally efficient solution for approximating the posterior distribution and replacing the continuously indexed Gaussian process (GP) with a discretely indexed Gaussian Markov random field (GMRF; Rue and Held, 2005). The INLA-SPDE approach is well known to be fast, accurate and efficient (e.g., de Rivera et al. 2018; Sadykova et al. 2016; Cosandey-Godin et al. 2015; Pinto et al. 2019; Quiroz et. al), and reduces the cost of computation from order $O(n^3)$ to order $O(n^{\frac{3}{2}})$ and $O(n^2)$ for the sparse spatial and spatio-temporal GMRFs, respectively (Cameletti et al. 2012; Rue & Held, 2005; Lindgren et al. 2011). Yet the application of such powerful methods in investigating spatio-temporal dynamics using joint models within the context of marine ecology is still underdeveloped. Here, we use the case study of herring, sprat and mackerel which feature in multiple acoustic and bottom trawl surveys in the North Sea to i) demonstrate the applicability of INLA-SPDE to combine datasets originating from different sources (surveys) and using different methods (e.g. bottom trawl and fisheries acoustics) and ii) provide information useful for fish stock assessment.

To the best of our knowledge, this study presents the first attempt to combine multispecies acoustic-trawl and trawl-trawl data utilising Bayesian hierarchical hurdle modelling approach, within the INLA-SPDE framework while simultaneously accounting for gear efficiencies, excess zeros and spatio-temporal autocorrelations within the data. Specifically, we assess whether, relative to individual survey estimates, models which combine data provide: i) more precise and stable estimates, ii) contrasting spatio-temporal biomass distributions, and iii) contrasting ecosystem-level temporal trends. We anticipate that our statistical approach will provide a useful methodological development capable of integrating a broad range of observations typical in ecology where diverse interests, from conservation to food security, are available for many species within an ecosystem.

## Material and Methods

### Data description

Pelagic fish data were obtained from several different sources: the North Sea component of the annual summer Herring Acoustic Survey (HERAS, ICES, 2020); the Quarter 3 (August – September), the International Bottom Trawl Survey (IBTS) and the Dutch Beam Trawl Surveys (BTS). HERAS is an ICES-coordinated annual summer acoustic survey specifically designed to assess herring stocks but also sprat (HERAS, ICES, 2020). The IBTS is internationally coordinated with records starting in the early 1980s and uses the Grande Overture Vertical (GOV) otter trawl with a vertical opening between 4 – 5m towed near the seabed to target demersal fish. The survey uses a fixed station sampling design with one sampling location per ICES rectangle per annual quarter, covering much of the North Sea (Fig 1). The Dutch BTS, first started in 1988, uses 8m wide beam trawls at the seabed in the third quarter (Q3; typically August or September) to survey flatfish stocks, e.g. sole and plaice, with hauls mostly aggregated in the southern North Sea (Fig 1; ICES, 2009, 2012). Pelagic species are not the target fish of the IBTS or BTS, but herring, sprat and mackerel are regularly observed in catches. All trawl data are available from DATRAS (https://datras.ices.dk/).

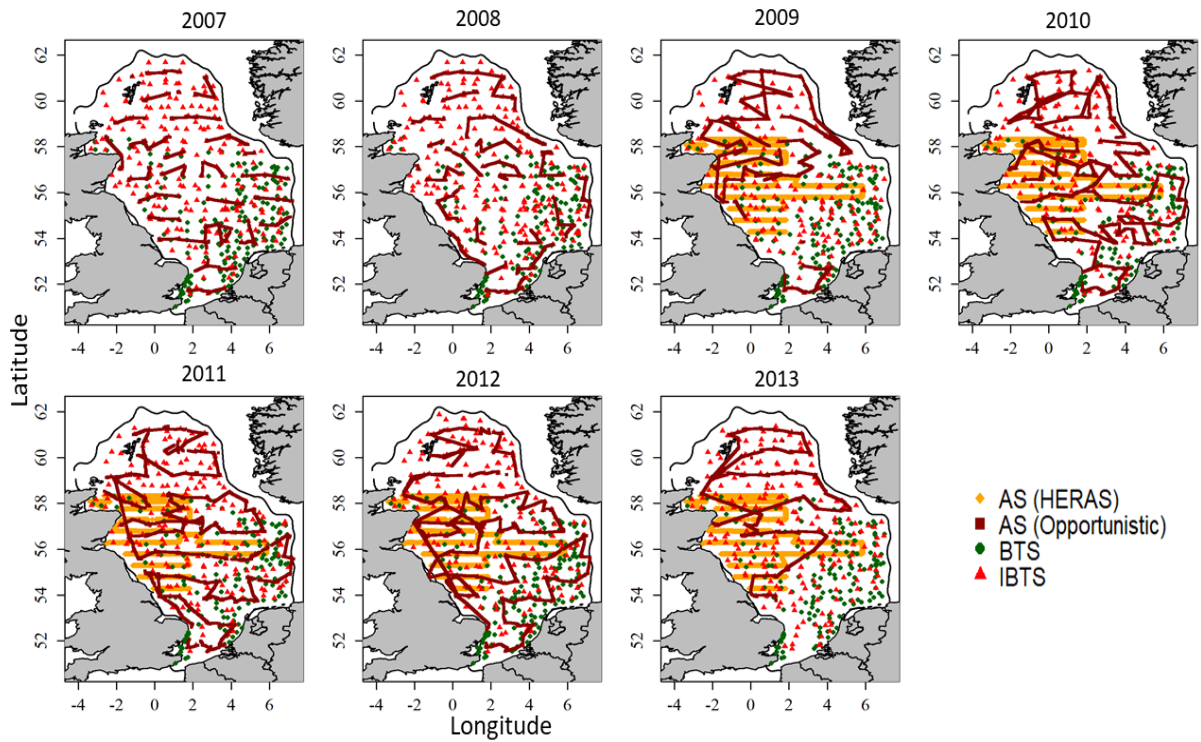

**Fig 1**. Distribution of transect tracks of the acoustic surveys and sampling locations for the trawl surveys. (AS- Acoustic Survey; BTS- Beam Trawl Survey; IBTS- International Bottom Trawl Survey) from 2007 to 2013. Note that AS (HERAS) data used here covered from 2009 – 2013 only. Brief descriptions of the different survey designs are given in Table 1 along with the sample sizes.

We used data from AS opportunistically collected aboard the *RV Cefas Endeavour* during the English component of the Q3 IBTS from 2007 to 2013 for mackerel, with three split-beam transducers of frequencies of 38kHz, 120kHz and 200kHz. The transducers were deployed at approximately 2m below the haul (7.7 m below the surface) with one ping per second. Fig 1 shows the transect tracks of the acoustic survey (brown tracks) with most of the northern North Sea covered and with almost no observations in the south in 2009 and 2013. We used the corresponding nautical area scattering coefficient (NASC in $m^2 nm^{-2}$) as a proxy for biomass (see, van der Kooij et al. 2015, for more details).

In addition, we used data from AS dedicated for herring and sprat carried out in the North Sea in Q3 and covering 2009 through to 2013.

In all, we extracted twelve datasets from the Q3 AS, BTS and IBTS spanning 2007 to 2013 for mackerel and 2009 to 2013 for herring and sprat to validate our approach (Table 1).

Table 1: Description of the gears/data and sample sizes across the three species for the 12 datasets explored in this study.

| Gear code | Description | Sample size ($n$) | | |
|---|---|---|---|---|
| | | Mackerel | Herring | Sprat |
| **IBTS** | Q3 International Bottom Trawl Survey with Grande Overture Verticale (GOV) otter trawl gear. | 1771 | 1235 | 1235 |
| **BTS** | Beam Trawl Survey with 8m wide beam trawl gear. | 1179 | 829 | 829 |
| **AS (HERAS)** | North Sea component of the summer herring acoustic survey, using acoustic methods with pelagic trawls | | 10998 | 10998 |
| **AS (opportunistic)** | Acoustic data collected opportunistically during English Component of the Q3 North Sea IBTS | 21599 | | |
| **Combined (IBTS + BTS + AS)** | Combination of the three principal surveys- IBTS, BTS and AS. | 24549 | 13062 | 13062 |

Note. Data on mackerel were for 2007 through to 2013, while data for herring and sprat covered 2009 through to 2013.

## Model specification

Let $y_s^{obs}$ be a generic term denoting the value of the response variable (e.g., biomass, NASC or abundance) with respect to a given specie in location $s$ ($s = 1, ..., n$). We focus on when observations are replicated over the $n$ spatial locations $T \geq 2$ times (or survey/sampling years) in which case we have $y_s^{obs} = y_{st}^{obs}$ $t$ ($t = 1, ..., T$). The response is assumed to come from a probability distribution such that $y_{st}^{obs} \sim \pi(\theta_{st})$ where $\theta_{st}$ is a generic term denoting the parameters of the probability distribution. Also, the probability density (or mass) function $\pi(.)$ could be continuous (e.g., Gaussian, gamma) or discrete (e.g., Poisson, binomial, negative binomial) depending on whether data are continuous or discrete counts. For example, in our continuous response case, we have that $\log(y_{st}^{obs}) \sim Normal(\theta_{st}, \sigma_e^2)$ where $\theta_{st}$ is the mean and $\sigma_e^2 > 0$ is the variance.

When data are zero-inflated and there is need to account for the zero and non-zero values to eliminate the possibility of predicting species as present when they have mean abundance of zero, a hurdle modelling technique can be used to model the zero and non-zero values separately by sub-setting the data into detection (or presence) and abundance (or biomass) which are assumed separate and orthogonal. Specifically, we employ the hurdle modelling approach and define a *detection* variable $z_{st}$ as

$$z_{st} = \begin{cases} 1, & \text{if } y_{st}^{obs} > 0 \\ 0, & \text{otherwise} \end{cases} \qquad eqn\ 1$$

while the *abundance* variable $y_{st}$ is defined as

$$y_{st} = \begin{cases} NA, & \text{if } y_{st}^{obs} = 0 \\ y_{st}^{obs}, & \text{otherwise} \end{cases}. \qquad eqn\ 2$$

Then, it is natural to assume that the binary detection variable Bernoulli probability mass function , so that, $z_{st} \sim Bernoulli(p_{st})$, where $p_{st}$, is the probability of detection of a specie in location $s$ in survey year $t$. Note that $p_{st}$ is also called gear efficiency, that is, the probability of catching and retaining fish found in the path of a gear and which depends on gear and fish characteristics including gear modifications, fish size and behaviour (e.g., Walker et al. 2017, Zhou et al. 2014, Fraser et al. 2007). If $z_{st} = 1$, that is, the specie is detected, the abundance variable is allowed to come from a continuous distribution that is positively skewed and heavy tailed such as the lognormal or gamma probability distribution. Note that when data are on a discrete count scale, a suitable discrete probability mass function such as the Poisson, Binomial or negative binomial may be assumed for the nonzero data. Nevertheless, for our purposes, as highlighted above, the logarithm of the abundance variable is assigned a normal distribution such that, $\log(y_{st}) \sim Normal(\mu_{st}, \sigma_e^2)$, where $\mu_{st} > 0$ is the mean and $\sigma_e^2$, from its origins in geostatistics, is also called the *nugget* effect (e.g., Cressie 1993, Camelleti et 2012; Krainski et al., 2019), and is the scale or variance parameter.

Then, estimates of $\theta_{st}$ are obtained as the product of expected biomass $\mu_{st}$ and the detection probability, that is, $\theta_{st} = \mu_{st} p_{st}$. Thus, the marginal likelihood of the data $y^{obs}$ given the unknown parameters $(p, \mu, \sigma_e^2)$ is given by

$$L(y|p, \mu, \sigma_e) = \prod_{s,t \in \mathcal{J}} \{\pi(y_{st}|\mu_{st}, \sigma_e) p_{st} I_{\{y_{st}>0\}} + (1 - p_{st}) I_{\{y_{st}=0\}}\} \qquad eqn\ 3$$

where $\mathcal{J}$ represents the index sets for $(s, t)$ and $I_x$ is an indicator function. To incorporate the various sources of variability and spatio-temporal autocorrelation structure inherent within the data, we assume that the data depend on a set of latent covariates through its mean function linked via the *additive* predictors $\eta^{(1)}$ and $\eta^{(2)}$ with appropriate link functions. Our hierarchical models are thus specified as follows:

**Observation**: $\pi(y_{st}|p_{st}, \mu_{st}, \sigma) = \pi(y_{st}|\mu_{st}, \sigma) p_{st} I_{\{y_{st}>0\}} + (1 - p_{st}) I_{\{y_{st}=0\}}$

**Process**: $logit(p_{st}) = \eta^{(1)} = \sum_{i=1}^{L} \beta_i^{(1)} specie_i + \sum_{j=1}^{K} f_j^{(1)}(gear) + f_{s,t}^{(1)}(location \times year)$

$$log(\mu_{st}) = \eta^{(2)} = \sum_{i=1}^{L} \beta_i^{(2)} specie_i + \sum_{j=1}^{K} f_j^{(2)}(gear) + f_{s,t}^{(2)}(location \times year)$$

$$f_{s,t}^{(.)}(.) = \rho f_{s,t-1}^{(.)}(.) + \omega_{s,t} \qquad eqn\ 4$$

Note that in equation (4), we have exploited the assumed conditional independence that exists between the data and the spatio-temporal random effects (e.g., Rue et al. 2009), where $logit(.)$ and $log(.)$ are the link functions connecting $\eta^{(1)}$ and $\eta^{(2)}$ to the probability of detection and mean biomass (or abundance), respectively; $specie_i$ is a dummy variable that takes the value 1 when the $i^{th}$ specie is considered and 0 otherwise with corresponding fixed effects coefficients $\beta_1, \ldots, B_L$; $f_j^{(.)}(.)$ is a smooth function for the gear-specific zero mean Gaussian random effects that capture the variability in the data due to the $K$ data sources; $f_{s,t}^{(.)}(.)$ is a continuously indexed Gaussian field (GF) that changes in time with first order autoregressive (AR1) dynamics and zero mean Gaussian innovations $\omega_{s,t}$ (Camelletti et al 2012) and which represents the spatio-temporal random effects for the interaction in space and time ($location \times year$), with $t = 2, \ldots, T$ and $|\rho| < 1$. We assume that the spatial process evolves temporally following autoregressive dynamics (Harvill, 2010), so that the spatio-temporal covariance matrix $\Sigma = \Sigma_s \otimes \Sigma_t$ (where $\otimes$ is a Kronecker product) is specified with a purely spatial covariance matrix. This implies that $\Sigma_{ss'} = \sigma_\omega^2 C(\Delta_{ss'})$, where $\sigma_\omega^2$ ($\sigma_z^2$ for the detection model or $\sigma_y^2$ for abundance model) is the marginal variance of the process, and $C(\Delta_{ss'})$ is a distance dependent Matérn correlation function given by

$$C(\Delta_{ss'}) = \frac{2^{1-\nu}}{\Gamma(\nu)} (\kappa \Delta_{ss'})^\nu K_\nu(\kappa \Delta_{ss'}) \qquad eqn\ 5$$

where $\Delta_{ss'} = ||s - s'||$ is the Euclidean distance between locations $s$ and $s'$; $\kappa = \sqrt{8\nu}/r$ is a scaling parameter and the range $r$ is the distance at which spatial correlation 0.13 for each smoothness parameter $\nu$ (Lindgren et al, 2011). Moreover, $K_\nu$ is the modified Bessel function of the second kind with order $\nu > 0$ (Abramowitz and Stegun, 1972).

*Bayesian inference via INLA-SPDE framework*

Based on the conditional independence assumption with Markovian dynamics, we used the R-INLA package (http://www.r-inla.org; Rue et al., 2009, Lindgren & Rue 2015) to evaluate the joint posterior distribution of the vector of the latent variable $\boldsymbol{\theta} = (\{f_j^{(.)}\}, \{f_{s,t}^{(.)}\})$ and the hyper-parameters $\boldsymbol{\psi} = (\sigma_e, \sigma_f, \sigma_\omega, \rho, \kappa)$ given all observations measured at time $t$, $\boldsymbol{y}_t = (y_{1,t}, \ldots, y_{n,t})'$ and approximated by

$$\pi(\boldsymbol{\theta}) \times \left(\prod_{t=1}^{T} \pi(\boldsymbol{y}_t|\psi_t, \boldsymbol{\theta})\right) \times \left(\pi(\psi_1|\theta) \prod_{t=2}^{T} \pi(\psi_t|\psi_{t-1}, \theta)\right) \qquad eqn\ 6$$

where $\pi(\boldsymbol{\theta})$ is a generic term representing the joint prior distribution of the parameters. (R Core Team, 2019). In R-INLA, robust and efficient numerical approximation of the marginal posterior distributions of the parameters are produced using Laplace approximation techniques (Martins et al. 2013). We used INLA in conjunction with the SPDE approach (Lindgren et al.). The SPDE approach uses basis function representation to define the continuously indexed GF with a discretely indexed Gaussian Markov Random field (GMRF; Rue and Held, 2005) built on the triangulation (mesh) of the study domain (entire North Sea). Instead, we specify the purely spatial random effect $f_{s,t} = f_s$ with

$$f_s = \sum_{m=1}^{M} B_m(s) f_m \qquad eqn\ 7$$

for a triangulation with $M$ vertices (nodes); where $\{B_m\}$ is the set of piecewise linear basis functions such that $B_m$ = 1 at vertex $m$, and 0 elsewhere; $\{f_m\}$ are zero mean Gaussian weights with sparse precision matrix $\boldsymbol{Q}_f$ which depends on the parameters of the Matérn correlation function defined in equation 5 and the smoothness parameter $\nu = 1$ (Blangiardo and Camelleti, 2013). Then, we replace the spatio-temporal random effect in equation 4 with $\sum_{m=1}^{M} \boldsymbol{A}_m f$, where $\boldsymbol{A}$ is an $n \times M$ sparse observation matrix that maps the GMRF $f$ from the $n$ observation locations to the $M$ nodes of the mesh (Blangiardo et al, Cameletti, Krainski et al). The mesh ($M = 399$ nodes) utilised for this study is shown in Fig S2 (supplementary material).

Furthermore, to arrive to a best-fit model, we carried out multiple adjustments to the models in equation 4 and tested for unadjusted effects of gear, space and time by running the models as described above but with the spatio-temporal random effect $f_{s,t}$ set to 0, $f_s$ and $f_t$, respectively. Note that the gear random effect $f_j$ was not included in the model for single survey datasets. For full Bayesian inference, we assigned the following default priors to the hyperparameters: $\tau_b \sim$

$logGamma(1, 0.00005)$; $\log(\kappa) \sim N(0,1)$; $\log\left(\frac{1+\rho}{1+\rho}\right) \sim N(0, 0.45)$. Models with the lowest Watanabe Akaike information criterion (WAIC) were considered to have the best fit (Watanabe, 2010). In all cases, the models which accounted for spatio-temporal autocorrelations provided the best fit (Table S1), hence we only report results from these models below. Species' relative biomass based on a given gear in year $t$, $b_t$, is calculated using the back transformed values of the additive predictors defined in equation (4) and averaged over survey year $t$, $\eta_t^{(1)}$ and $\eta_t^{(2)}$, such that

$$b_t = \left\{ \frac{\exp\left(\eta_t^{(1)}\right)}{1 + \exp\left(\eta_t^{(1)}\right)} \right\} \times \exp\left(\eta_t^{(2)}\right) \qquad eqn\ 8$$

Then, to compare temporal trends over the various survey years and across the different gears as well as the combined gears data, we calculate the scaled marginal posterior biomass estimate $\tilde{b}_t$ given by

$$\tilde{b}_t = \frac{b_t}{RMS_B} \qquad eqn\ 9$$

where $B = (b_1, b_2, \ldots, b_T)$ is the vector of specie's average biomass based on a given gear type over $T$ years. The term $RMS_B$ is the estimate of the *root-mean-square* based on the vector $B$ which is calculated as

$$RMS_B = \sqrt{\frac{\sum_{t=1}^{T} b_t^2}{T-1}} \ . \qquad eqn\ 10$$

In summary, we expect that our models which account for gear random effects (or gear efficiency) to combine disparate survey information will provide precise and more stable biomass estimates, with comparatively low standard deviation, relative to models based on individual surveys due to increased sample size and decreased credible interval width. This will provide a more detailed multispecies assessment of the biomass distribution across an ecosystem that could be readily extended to incorporate other species, including flatfish (best surveyed using BTS, but observed in IBTS catches) across various temporal and spatial scales.

## Results

Overall, our models that drew on the combined surveys (Combined or IBTS + BTS + AS) consistently provided more precise and more stable estimates of biomass distribution and detection probability relative to their unique survey counterparts (Fig 2, Table 2). Estimates of uncertainties in estimation based on the different datasets across all species show that the combined dataset provided overall most precise and stable estimates for both detection probability and abundance estimation (Fig 2). This was largely due to the increased spatial coverage (sample size) which in turn narrowed the width of the credible interval (CI) of posterior marginal parameter estimates and ultimately increased statistical power.

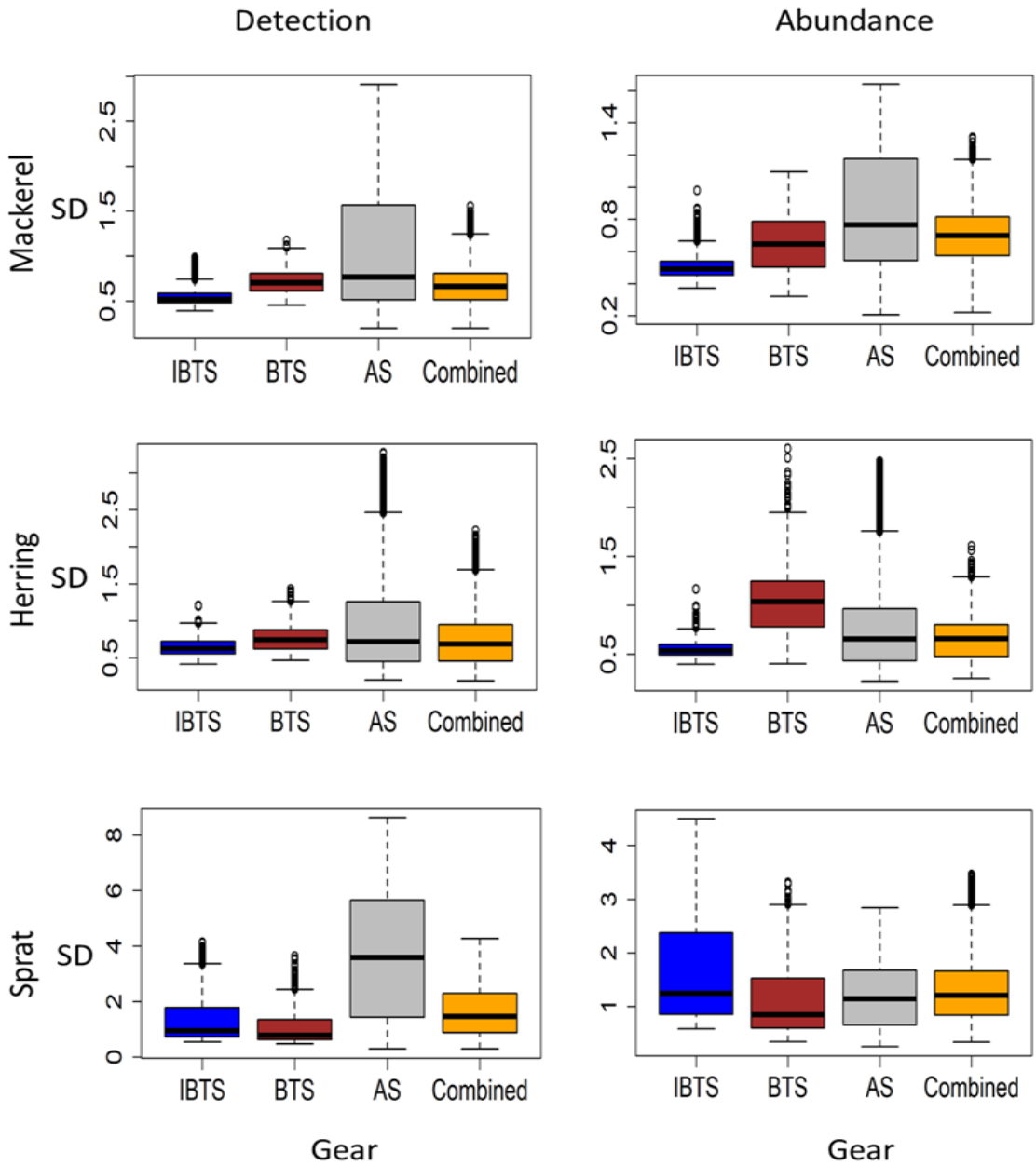

**Fig 2**. Posterior marginal estimates of the standard deviation for the estimation of detection probability and the abundance of the pelagic species based on the different datasets. AS- Acoustic Survey; BTS- Beam Trawl Survey; IBTS- International Bottom Trawl Survey; Combined (AS + BTS + IBTS, that is, all three principal surveys combined).

Table 2: Posterior marginal estimates of model parameters based on the combined data models for both detection and abundance models.

| Factor/variable | Mackerel (AS + BTS + IBTS) | | Herring (AS + BTS + IBTS) | | Sprat (AS + BTS + IBTS) | |
|---|---|---|---|---|---|---|
| | Mean | 95% CI | Mean | 95% CI | Mean | 95% CI |
| *Detection* | | | | | | |
| Marginal variance, $\sigma_z^2$ | 3.47 | (2.84, 4.13) | 6.98 | (5.55, 8.49) | 22.35 | (14.69, 30.50) |
| Interannual correlation, $\rho$(%) | 53.73 | (43.01, 62.99) | 41.17 | (28.79, 53.14) | 73.57 | (63.30, 82.28) |
| *Gear efficiency (%):* | | | | | | |
| AS | 4.07 | (2.85, 5.79) | 5.01 | (2.85, 8.66) | 0.09 | (0.01, 0.82) |
| BTS | 1.94 | (1.15, 3.20) | 1.59 | (0.80, 3.10) | 0.01 | (0.00, 0.12) |
| IBTS | 72.63 | (64.68, 79.39) | 88.75 | (81.8, 93.44) | 7.23 | (0.74, 40.11) |
| *Abundance* | | | | | | |
| Marginal variance, $\sigma_y^2$ | 1.55 | (1.20, 1.94) | 2.74 | (2.02, 3.53) | 7.80 | (3.28, 13.70) |
| Interannual correlation, $\rho$(%) | 52.2 | (34.27, 67.28) | 60.96 | (45.92, 74.68) | 87.95 | (72.87, 96.66) |
| *Gear effects:* | | | | | | |
| AS | 1.70 | (1.49, 1.92) | 2.88 | (2.12, 3.59) | 3.76 | (0.58, 5.89) |
| BTS | 5.20 | (4.52, 5.89) | 3.33 | (2.34, 4.30) | -0.20 | (-3.41, 2.02) |
| IBTS | 8.57 | (8.36, 8.77) | 8.32 | (7.57, 9.00) | 6.10 | (2.93, 8.19) |

CI: Credible Interval.

The spatio-temporal distribution of mackerel biomass varied across datasets (Fig. 3, Fig S4) with aggregations towards western and southern coastlines for the IBTS, but more northerly for AS, although both showed lower probability of detection in 2009, 2010 and 2013. The probability of mackerel detection was low at $p \leq 0.25$ across years for the BTS-only data. The spatio-temporal models based on the combined data drew largely on the IBTS and AS observations, showing biomass largely aggregated in the north, some to the west, central and south, with lower detection probability to the east, particularly in 2009, 2010 and 2013.

IBTS provided the least uncertainty (i.e. as defined by standard deviation) when estimating the detection probability of mackerel, while AS had relatively high uncertainty in the south and this was evident in the combined data assessment where uncertainty was generally low but higher in the south (Fig 3b).

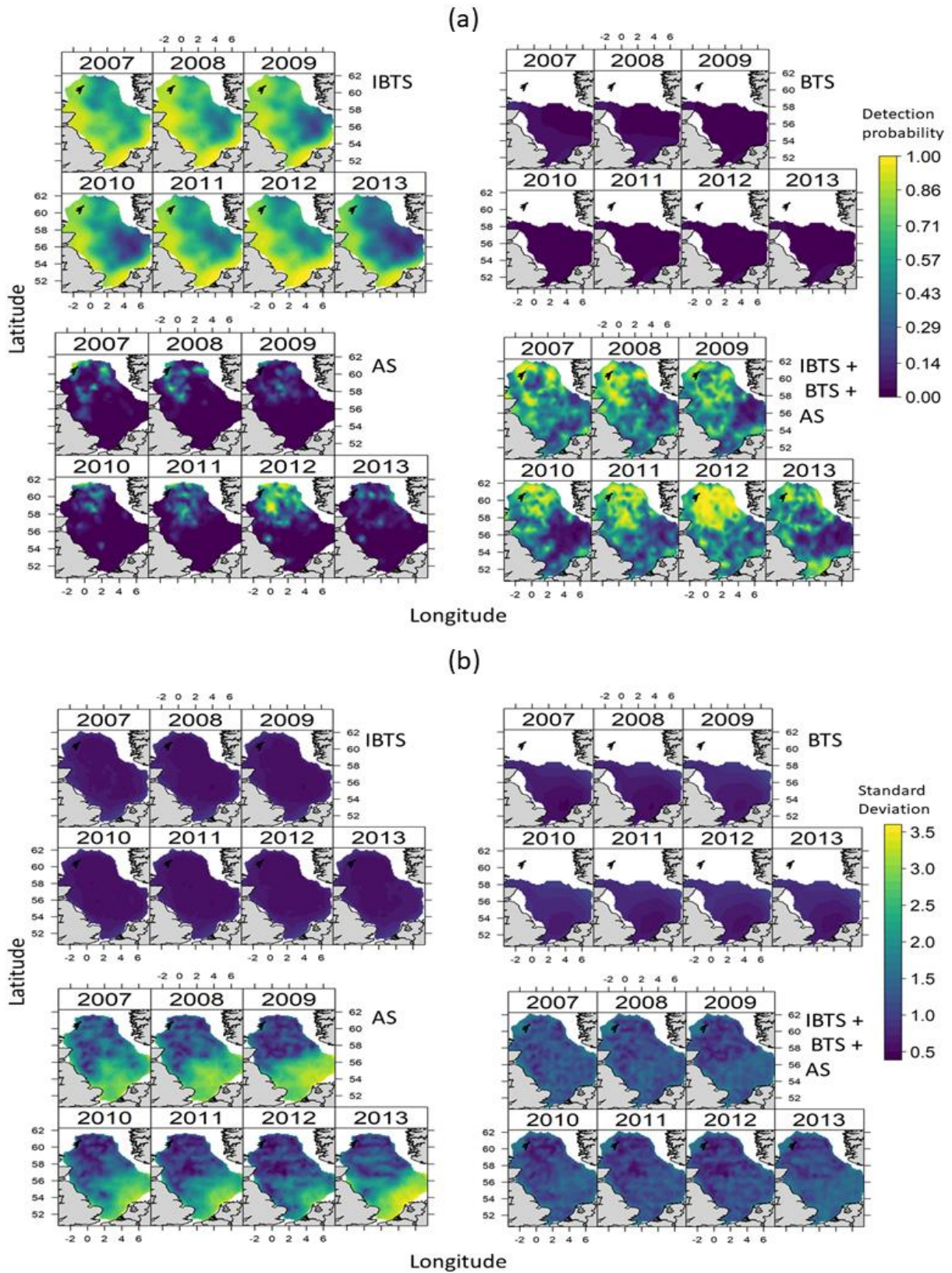

**Fig 3. (a)** Posterior marginal mean probability of detection of *mackerel* and **(b)** standard deviation (SD), from 2007 to 2013. AS- Acoustic Survey; BTS- Beam Trawl Survey; IBTS- International Bottom Trawl Survey; IBTS+BTS+AS- all three principal surveys combined. In these maps, dark blue to yellow correspond to locations with lowest to highest values.

The spatio-temporal distribution of herring was similar across all years with biomass aggregated in the north, west and south east based on the IBTS-only data, whereas the AS data showed greater interannual variability in the area that it covered (Fig 4, Supplementary material Fig S5). The probability of herring detection was low across years for the BTS-only data. The combined data showed broadly similar patterns in the spatio-temporal distribution of herring as the IBTS-data but with more interannual variability (Fig 4a). Among the individual surveys, the IBTS generally provided the most precise estimates and AS the least. Where data were combined, the regions not covered by the AS tended to have higher uncertainties, but these were reduced relative to the AS (Fig 4b, Fig S5).

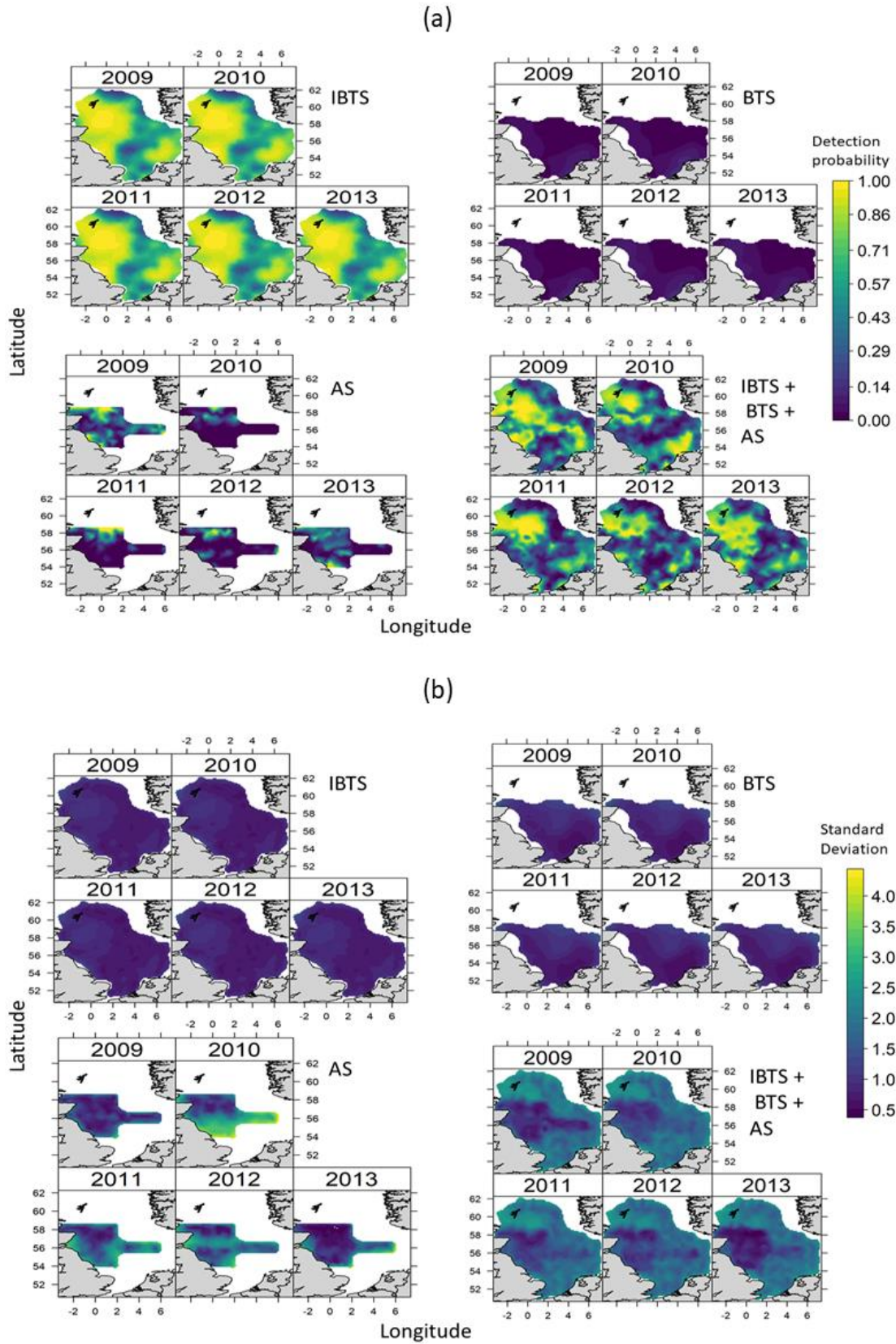

**Fig 4. (a)** Posterior marginal mean probability of detection of *herring* and **(b)** standard deviation (SD), from 2007 to 2013. AS- Acoustic Survey; BTS- Beam Trawl Survey; IBTS- International Bottom Trawl Survey; IBTS+BTS+AS- all three principal surveys combined. In these maps, dark blue to yellow correspond to locations with lowest to highest values.

Sprat were largely aggregated along coastlines to the west, south and south east (Fig. 5, S6) with limited interannual variability based on the IBTS-only data. This pattern that was generally supported by the AS and BTS data in the areas that they covered. Sprat were marginally more likely to be caught than mackerel or herring in the BTS (Fig 5a). The combined data showed broadly similar patterns in the spatio-temporal distribution of sprat as the IBTS-data with relatively higher uncertainties in areas without AS data. Similar patterns in spatio-temporal trends as well as the standard deviations were obtained for the posterior marginal mean of the spatial random effects for each species abundance as well as the corresponding standard deviation (Figs S4-S6).

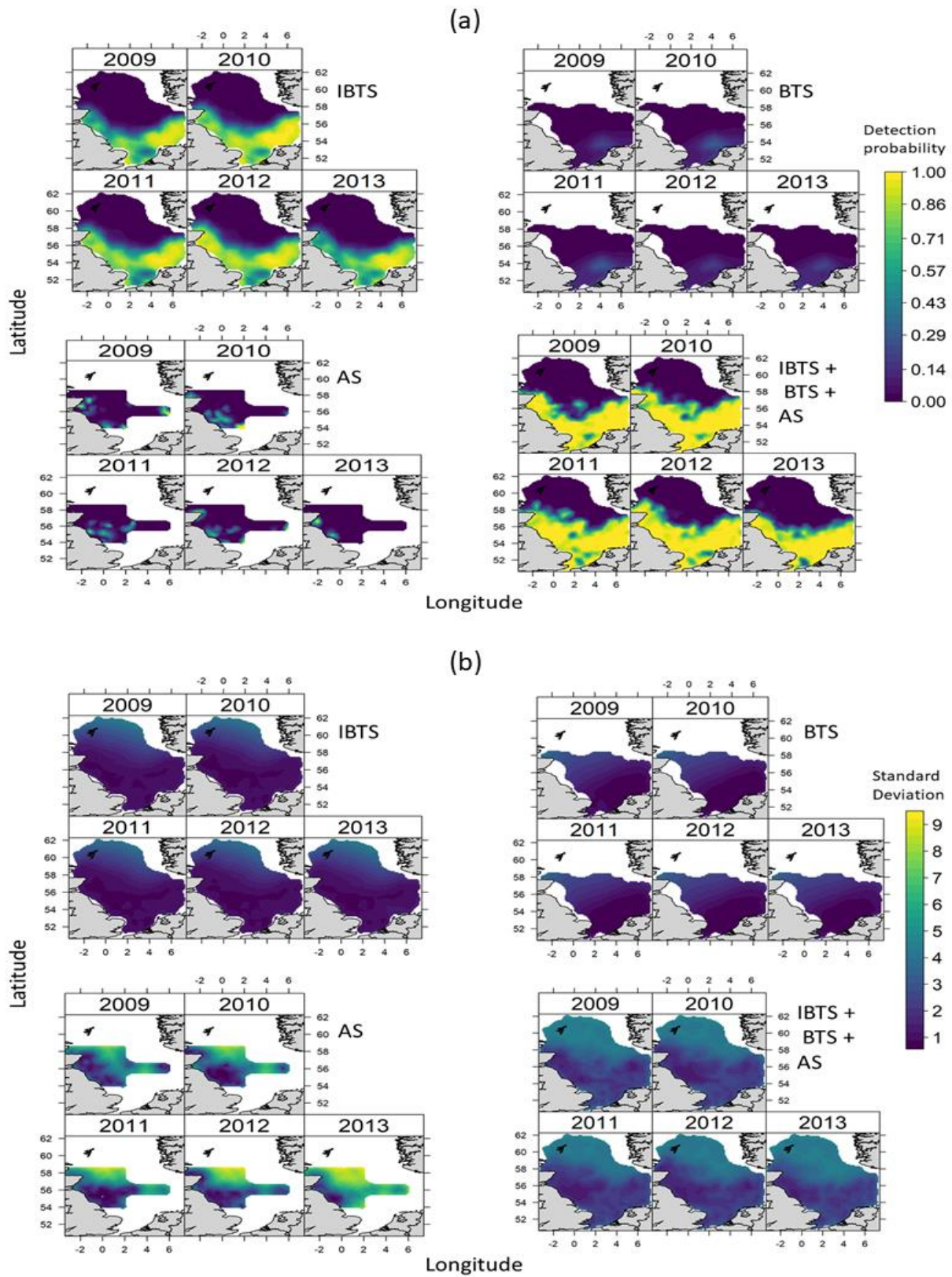

**Fig 5. (a)** Posterior marginal mean probability of detection of *Sprat* and **(b)** standard deviation (SD), from 2007 to 2013. AS- Acoustic Survey; BTS- Beam Trawl Survey; IBTS- International Bottom Trawl Survey; IBTS+BTS+AS- all three principal surveys combined. In these maps, dark blue to yellow correspond to locations with lowest to highest values.

Temporal North Sea-scale relative biomass trends in the combined data show that mackerel had broadly similar biomass in 2007 and 2013 with a low point in 2010, while the trawl surveys and AS show opposing decreasing and increasing trends, respectively (Fig. 6). North Sea-scale relative biomass trends for herring show a highpoint in 2010 but an overall decrease between 2009-2013 based on the combined data, the AS data also shows a decreasing trend over the time-series, whereas trawls show a slight increasing trend. For sprat, North Sea-scale relative biomass trends were similar across the combined and IBTS data, declining between 2009-2013, AS data also showed a decline but with much more interannual variability, and BTS had high variability but no clear change overall.

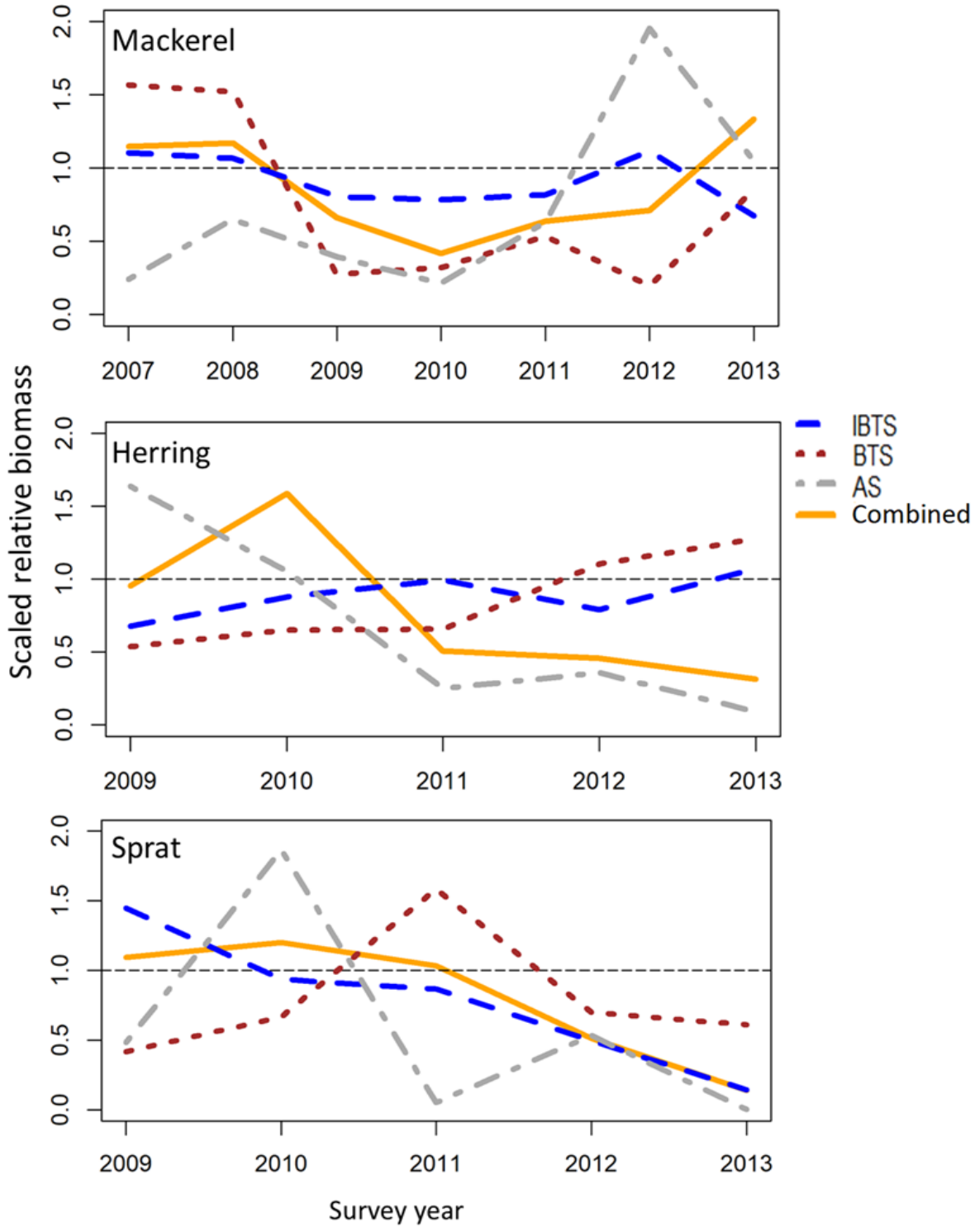

**Fig 6.** Comparison of the scaled abundance indices derived from the back transformed results of the spatio-temporal models fitted to the 8 datasets from the three principal surveys (AS, BTS, IBTS) across the three species (mackerel, herring & sprat). AS- Acoustic Survey; BTS- Beam Trawl Survey; IBTS- International Bottom Trawl Survey; Combined (or IBTS+BTS+AS)- all three principal surveys combined.

## Discussion

Methods that can integrate data on multiple species from different surveys and sampling tools will help to improve stock assessments and indeed aspirations for ecosystem-based fisheries management. Previously, differences in gear efficiency have only been measured using multiple trawl surveys (e.g., Moriarity *et al.,* 2020; Walker *et al* 2017). We extend this to include multiple trawl surveys with an acoustic survey within a fully Bayesian framework that incorporates multiple sources of uncertainties to estimate both the probability of detecting schooling fish and their biomass. Specifically, we could predict the probability that an individual fish or school of fish would be encountered and thus get closer to the 'true' biomass distribution for these species across the North Sea, including in areas where data existed for only one survey. We also found that data from differing sources often provided contradictory temporal trends in species stocks. Thus, by accounting for uncertainties unique to discrete data sources, we were able to achieve the most representative assessment of the fish true population status from the combined information. This approach therefore attempts to address the recommendation following a global assessment of fishing impacts on marine biodiversity (Branch et al. 2010) that true population trends for marine species, particularly those most susceptible to fishing pressure, are better assessed.

There has been much recent progress in the combination of data from across surveys, such as the application of generalized additive models (GAMs; Walker et al., 2017), generalized additive mixed models (GAMMs; Moriarty et al. 2020), beta regression MCMC (Kotiwicki et al., 2017; Juntunen et al 2012), and INLA (Pinto et al. 2019). These advanced statistical approaches have their drawbacks. For eample, GAMs and GAMMs are non-Bayesian and the quantification of uncertainties in parameter estimates are generally not straightforward .

Also, approaches that relied on simulation based techniques such as beta regression MCMC are known to deteriorate rapidly as the data dimension increases (Quiroz et al. 2015; Sadykova, 2019), and studies which simultaneously accounted for variations in gear efficiencies (effects) and zero-inflated data within the INLA context are generally lacking. Here, we present binomial-hurdle model implemented via INLA-SPDE specifically to address these issues. The approach allowed us to combine data from disparate sources over same or different temporal and/or spatial scales thereby enabling us to gain increased spatial coverage of our assessment, giving rise to more precise estimates than the individual surveys. Although for the illustrative data utilised here, we note that the IBTS provided the most precise estimates among the individual surveys across all the three species, however, the combined data provided the more stable estimates in general (Fig 2 & Fig 6).

Varying results across gears, e.g., higher biomass of mackerel detected in the north using AS but more widely distributed for the trawls, suggests different gears detect different behavioural components. While acoustic methods are better at detecting schooling mackerel, they are less able to detect dispersed fish, particularly in the case of mackerel which, due to the lack of a swim bladder, has a relatively weak acoustic signal (van der Kooij et al., 2016). During the summer, a component of the mackerel population appears to disaggregate to feed and, in the southern North Sea, juvenile mackerel appear to be more dispersed in the water column. Thus, it makes sense to use a trawl in some areas and acoustic methods in others, with intersecting areas. In our case with mackerel, the 'intersection' was provided by opportunistically collected acoustic observations between trawls. This opens the door for ecosystem monitoring designs that are not reliant only on one standardised approach, such as IBTS, but multiple partly overlapping surveys using specific gear for particular habitats (e.g. shallow coastal vs offshore) or to capture specific behaviour (e.g. schooling). Our approach means this is now feasible within a single statistical framework, drawing on the strengths of multiple disparate surveys to provide an ecosystem-wide multispecies assessment.

Rather than an exhaustive exercise in analysing all available North Sea fish survey data, our work should be viewed as a methodological marker highlighting that future fish population assessments could include data from across surveys and species. We anticipate future work to include species-size-classes relevant to stock assessment. For example, Flatfish which are best surveyed using the BTS but are observed in the IBTS and draw on the wealth of other acoustic and trawl data, including those from industry, across species available for the North Sea. By adapting our methodological demonstration, the approach could be readily applied to many fish species and marine ecosystems where disparate surveys intersect, such as trawl-trawl, trawl-acoustic or some more complex combination. Such an approach is also not restricted to marine fisheries, but applicable to assess population and biodiversity trends where there are many different types of observations, including those more opportunistic from citizen scientists alongside expert surveys (e.g. OBIS 2020; Global Biodiversity Information Facility; http://www.gbif.org/).

Our approach could be developed further by fitting Environmental, biotic and anthropogenic drivers such as depth, species interactions and fishing pressure as covariates in our models enabling future projections of multiple species biomass under a range of management scenarios, for example. This, coupled with estimates across many species, could allow predictions of change in ecosystem structure and function based on functional traits (e.g. using feeding guilds, Thompson et al. 2020) under varying environmental change scenarios. Developments such as these will be critical to develop a process-based understanding of ecosystem structure and function to better underpin ecosystem-based management.

# Acknowledgements

Cefas Seedcorn funding for DP427 'Forecasting and valuing changes in food web structure and function in response to environmental change'.# Conflict of Interest

Authors declare no conflict of interest.

# Author contributions

CCN, MSAT, MAS, EC, JvdK & CL conceived the ideas; CCN & MAS designed the methodology; CCN developed and implemented the methodology; MSAT secured the funding; CL collated the data; CCN analysed the data; CCN led the writing of the manuscript. All authors contributed critically to the drafts and gave final approval for publication.

# References

Abramowitz M and Stegun I. (1972). Handbook of mathematical functions. New York: Courier Dover Publications.

Blangiardo, M., Cameletti, M., Baio, G., and Rue, H. A (2013). Spatial and spatio-temporal models with R-INLA. *Spat Spatiotemporal Epidemiology*, 7:39-55.

Branch, T. A., Watson, R., Fulton, E. A., Jennings, S., McGilliard, C. R., Pablico, G. T., Ricard, D., & Tracey, S. R. (2010). The trophic fingerprint of marine fisheries. *Nature*, *468*(7322), 431–435. https://doi.org/10.1038/nature09528

Cameletti, M., Finn Lindgren, F., Simpson, D., and Rue, H. (2012). Spatio-temporal modeling of particulate matter concentration through the SPDE approach. Adv. Stat. Anal. **97**(2): 109–131. doi:10.1007/s10182-012-0196-3..

Cheung, W. W. L., Lam, V. W. Y., Sarmiento, J. L., Kearney, K., Watson, R., & Pauly, D. (2009). Projecting global marine biodiversity impacts under climate change scenarios. *Fish and Fisheries*. https://doi.org/10.1111/j.1467-2979.2008.00315.x29